\documentclass[10pt]{article}
\usepackage{subfigure}
\usepackage{fullpage}
\usepackage{setspace}
\usepackage{parskip}
\usepackage{titlesec}
\usepackage[section]{placeins}
\usepackage{xcolor}
\usepackage{breakcites}
\usepackage{lineno}
\usepackage{hyphenat}
\usepackage{footmisc}

\PassOptionsToPackage{hyphens}{url}
\usepackage[colorlinks = true,
            linkcolor = blue,
            urlcolor  = blue,
            citecolor = blue,
            anchorcolor = blue]{hyperref}
\usepackage{etoolbox}

\renewenvironment{abstract}
  {{\bfseries\noindent{\abstractname}\par\nobreak}\footnotesize}
  {\bigskip}

\titlespacing{\section}{0pt}{*3}{*1}
\titlespacing{\subsection}{0pt}{*2}{*0.5}
\titlespacing{\subsubsection}{0pt}{*1.5}{0pt}

\usepackage{authblk}

\usepackage{graphicx}
\usepackage[space]{grffile}
\usepackage{latexsym}
\usepackage{textcomp}
\usepackage{longtable}
\usepackage{tabulary}
\usepackage{booktabs,array,multirow}
\usepackage{amsfonts,amsmath,amssymb}
\providecommand\citet{\cite}
\providecommand\citep{\cite}

\newif\iflatexml\latexmlfalse
\providecommand{\tightlist}{\setlength{\itemsep}{0pt}\setlength{\parskip}{0pt}}%

\AtBeginDocument{\DeclareGraphicsExtensions{.pdf,.PDF,.eps,.EPS,.png,.PNG,.tif,.TIF,.jpg,.JPG,.jpeg,.JPEG}}

\usepackage[utf8]{inputenc}
\usepackage[english]{babel}

\usepackage{float}

\usepackage[margin=1.5in]{geometry}

\begin{document}

\title{Mining the online infosphere: A survey}

\author[1]{Sayantan Adak}
\author[1]{Souvic Chakraborty}
\author[1]{Paramtia Das}
\author[1]{Mithun Das}
\author[1]{Abhisek Dash}
\author[2]{Rima Hazra}
\author[1]{Binny Mathew}
\author[1]{Punyajoy Saha}
\author[1]{Soumya Sarkar}
\author[1]{Animesh Mukherjee}%

\affil[1]{Department of Computer Science and Engineering, Indian Institute of Technology, Kharagpur, West Bengal, India -- 721302}%
\affil[2]{Advanced Technology Development Center, Indian Institute of Technology, Kharagpur, West Bengal, India -- 721302\thanks{The first nine authors have been arranged based on family names and have equal contributions.}}
\vspace{-1em}

  \date{}

\begingroup
\let\center\flushleft
\let\endcenter\endflushleft
\maketitle
\endgroup

\selectlanguage{english}
\begin{abstract}
The evolution of AI-based system and applications had pervaded everyday life to make decisions that have momentous impact on individuals and society. With the staggering growth of online data, often termed as the \emph{Online Infosphere} it has become paramount to monitor the infosphere to ensure social good as the AI-based decisions are severely dependant on it. The goal of this survey is to provide a comprehensive review of some of the most important research areas related to infosphere, focusing on the technical challenges and potential solutions. The survey also outlines some of the important future directions. We begin by discussions focused on the collaborative systems that have emerged within the infosphere with a special thrust on Wikipedia. In the follow up we demonstrate how the infosphere has been instrumental in the growth of scientific citations and collaborations thus fuelling interdisciplinary research. Finally, we illustrate the issues related to the governance of the infosphere such as the tackling of the (a) rising hateful and abusive behaviour and (b) bias and discrimination in different online platforms and news reporting.  

% The abstract should be a concise (less than 250 words) description of
% the article and its implications. It should include all keywords
% associated with your article, as keywords increase its discoverability.
% Please do not include generic phrases such as ``This article discusses
% \ldots{}'' or ``Here we review,'' or references to other articles. Note:
% You will be required to copy this abstract into the submission system
% when uploading your article.%
\end{abstract}%

\sloppy

%\textbf{Remember that you are writing for an interdisciplinary audience.
%Please be sure to discuss interdisciplinary themes, issues, debates,
%etc. where appropriate.} Note that the WIREs are forums for review
%articles, rather than primary literature describing the results of
%original research.

%\section*{Article Title}

%{\label{350486}}

%The title should not exceed 20 words. Please be original and try to include keywords, especially before a colon if applicable, as they will increase the discoverability of your article.~\href{https://authorservices.wiley.com/author-resources/Journal-Authors/Prepare/writing-for-seo.html}{Tips on Search Engine Optimization}

%\section*{Article Type}

%{\label{925764}}

%The \href{http://wires.wiley.com/go/forauthors\#ArticleTypes}{Article Type} denotes the intended level of readership for your article. Pleaseselect one of the below article type options. An Editor may havementioned a specific Article Type in yourinvitation letter; if so, please let them know if you think a different Article Type better suitsyour topic.~

\if{0}\begin{itemize}
\tightlist
\item
  Opinion
\item
  Primer
\item
  Overview
\item
  Advanced Review
\item
  Focus Article
\item
  Software Focus
\end{itemize}
\fi
%\section*{Authors}

\if{0}{\label{290010}}

List each person's full name, ORCID iD, affiliation, email address, and
any conflicts of interest. Please use an asterisk (*) to indicate the
corresponding author.

The preferred (but optional) format for author names is First Name,
Middle Initial, Last Name.~~

The submitting author is required to provide
a~\href{https://authorservices.wiley.com/author-resources/Journal-Authors/Submission/orcid.html}{ORCID
iD}, and all other authors are encouraged to do so.~

Wiley requires that all authors disclose any potential conflicts of
interest. Any interest or relationship, financial or otherwise, that
might be perceived as influencing an author's objectivity is considered
a potential conflict of interest. The existence of a conflict of
interest does not preclude publication.
\fi

\if{0}\section*{Abstract}

{\label{468816}}

The abstract should be a concise (less than 250 words) description of
the article and its implications. It should include all keywords
associated with your article, as keywords increase its discoverability
(\href{https://authorservices.wiley.com/author-resources/Journal-Authors/Prepare/writing-for-seo.html}{Tips
on Search Engine Optimization}). Please try not to include generic
phrases such as ``This article discusses \ldots{}'' or ``Here we
review,'' or references to other articles. Note: You will be required to
copy this abstract into the submission system when uploading your
article.

Optional: If you would like to submit your abstract in an additional
language,~\href{http://wires.wiley.com/go/forauthors\#Resources}{read
more}.

\section*{Graphical/Visual Abstract and
Caption}

{\label{750712}}

Include an attractive full color image to go under the text abstract and
in the online Table of Contents.~\textbf{You will also need to upload
this as a separate file during submission.~}It may be a figure or panel
from the article or may be specifically designed as a visual summary.
While original images are preferred, if you need to look for a
thematically appropriate stock image, you can go
to~\href{http://pixabay.com/}{pixabay.com}~(not affiliated with Wiley)
to find a free stock image with a CC0 license. Another option you have
is to utilize professional illustrators with
Wiley's~\href{https://wileyeditingservices.com/en/article-preparation/graphical-abstract-design}{Graphical
Abstract Design service}.

Size: The minimum resolution is 300 dpi. Please keep the image as simple
as possible because it will be displayed in multiple sizes. Multiple
panels and text other than labels are strongly discouraged.

Caption: This is a narrative sentence to convey the article's essence
and wider implications to a non-specialist audience. The maximum length
is 50 words, but consider using 280 characters or less to facilitate
social media sharing, which can increase the discoverability of your
article.\fi

\par\null

\section{Introduction}

{\label{252565}}

\textit{Online infosphere}\footnote{\url{https://en.wikipedia.org/wiki/Infosphere}} is the term corresponding to the Internet becoming a virtual parallel world formed from billions of networks of \textit{artificial life} at different scales ranging from tiny pieces of software to massive AI tools running a factory or driving a car. The motivations for this are diverse, seeking to both help mankind and harm it.

In this article, we shall attempt to portray some of the areas that are increasingly gaining importance in research related to the evolution of this infosphere. In particular, we would begin with infosphere as a collaborative platform, Wikipedia being the prime point of discussion. As a next step we would discuss how the infosphere has influenced the evolution of scientific citations and collaborations. Finally, we shall outline the emerging research interest in the governance of this infosphere to eradicate discrimination, bias, abuse and hate.   

\subsection{Infosphere as a collaborative platform}
The infosphere hosts numerous collaborative platforms including question answering sites, folksonomies, microblogging sites and above all encyclopedias. In this survey we shall focus on Wikipedia which is one of the largest online collaborative encyclopedia. We shall primarily discuss two of the most important aspects of Wikipedia -- (a) the quality of an article and its indicators and (b) the collaboration dynamics of Wikipedia editors who constitute the backbone of this massive initiative. Under the first topic we shall identify the different features of an article like its language, structure and stability as well as their quality~\cite{zhang2020mining,halfaker2019ores,sarkar2019stre}. We shall further summarise attempts that have been made to automatically predict the quality of an article~\cite{guda2020nwqm,marrese2019edit}. Within the second topic we shall briefly describe various issues related to the community of editors including anomalies, vandalism and edit wars\cite{Kiesel2017SpatioTemporalAO,Tran2015CrossLanguageLF}. Finally, we shall talk about ways to enabling retention of editors on the platform~\cite{halfaker2013rise,muric2019collaboration,steinmacher2015social}.

\subsection{Infosphere shaping scientific citations and collaborations}%Not written  about collaboration networks etc in the section.
Citations play a crucial role in shaping the evolution of a scientific discipline. With an exponential growth of research publications in various disciplines it has become very important for researchers and scientists to grasp different concepts within a short period of time. We would explore how the infosphere has influenced the growth and interaction of different scientific disciplines over the period of last few years by investigating several different aspects of citation networks. Our survey includes -- (a) a detailed account of how the basic sciences and the computer sciences have interacted with each other over the years resulting in an interdisciplinary research landscape~\cite{Hazra:2019,Morillo:2003},%Here I explained mainly the usage of citation network in order observe the interdisciplinary research 
 (b) the temporal dynamics of citations~\cite{relaylink}, (c) ways for assessment of article quality, %Sir, after the "dynamics of citations" section. I have mentioned two sections (a) citation count prediction (of articles), (b) citation recommendation: In citation reco, when authors write some sentences in their articles, model will automatically recommend them best citation possible.
and finally (d) a brief account of anomalous citation flows.

\subsection{Governance of the infosphere}
The stupendous growth of the infosphere has resulted in the emergence of various online communities that have massively started infusing bias, discrimination, hatred and abuse often resulting in violence in the offline world. In this segment, we shall primarily focus our discussion on the following topics -- (a) analysis, spread, detection and mitigation of online hate speech and (b) biases that manifest across news media and in traditional recommendation systems. Within the first topic we shall motivate the need to tackle online hate speech by citing some of the adverse consequences of the same. In particular, we shall see how unmoderated hate speech spreads in a social network~\cite{Mathew2019,binny_temporalCSCW}, what are the challenges to automatically detect online hate speech~\cite{Gomez2020,aluru2020deep} and the possible techniques to combat this problem~\cite{mathew2020interaction,binnyCounter}. Within the second topic we shall discuss two important forms of biases. The first one corresponds to political biases that manifest due to the massive production of unverified (and in many cases false) news generated in the form of news/blog/tweets etc. We shall also discuss the difficulties faced by modern machine learning techniques in preventing the infusion of such biases. The second one narrates the idea of formation of \textit{filter bubbles}~\cite{pariser2011filter} in traditional recommendation systems followed by a discussion on the need to systematically audit such systems~\cite{dash2019network, dash2021When}. 

%Under the first section we will first elaborate the adverse impacts of hatespeech in a society by citing the recent relevant articles and additionally show our study on the spread of hatespeech in online social networks. Then we shall discuss the challenges detecting hateful posts in social media and important researches on the network of hateful users. Finally we will present the recent studies on automatically preventing the hatespeech. For the second section we shall describe the recent trends of political bias among the news media and spreading of unverified contents within the infosphere and the difficulties regarding the prevention using AI based systems. Finally we leverage the Filter Bubble problem in recommendation system and investigate the recent studies on improving the fairness in Recommendation Systems. 

\if{0}{
In this article we have leveraged our four recent and important study on hate-speech, bias and fairness, citation dynamics and wikipedia. 
With the exponential growth of using social media such as twitter, facebook etc the online hatespeech has become pervasive and raise a
serious concern for online communities, social media platforms and government organisations. At the extreme scenario this could even 
cause for genocide, riots and mass shooting. Though the actual definition of hate-speech varies according to the literature it can be 
generalized as "abusive language that is used to express hatred towards a targeted group or individual".  In the study we have focused on
the challenges detecting hate-speech automatically in online social media platforms, advantages and disadvantages previous research 
works, user-based approach for detecting hate-speech and it's spread.  We have analysed the network structure of the spread of 
hate-speech in online platforms. Lastly we have shown the study of Counter speech to tackle the online hate-speech automatically.

Secondly we have shown our study regarding media bias and fairness in recommendation systems. As the booming increase of usage 
of internet, more specifically social media the sharing of unverified content has become a trend for the normal people as well as for the 
news media and this has led to fake news propagation by the most of the reputed news media. These unverified contents mostly inclined 
towards particular aspect and quantifying the bias has become an important area for research. We have shown the vulnerabilities for a 
mathematical models for detecting media bias as it is strongly relies on the beliefs of the data annotators. Now coming to the recommendation
system, it is one of the most important features for any e-commerce sites to provide a strong relevant product information to a customer. 
However, due to the over dependency on relevance has led to a bias on the topics of information. We leverage the Filter Bubble problem in
this article and have shown our research on fairness in recommendation systems.

Detecting the citation dynamics and predicting the emerging topics is always an important field of research. We have framed a citation network
and tried to analyze the corresponding research problems with the help of citation network. We have shown the interdisciplinary research in terms of
citation interaction and observed the interdisciplinary citation flow. We also analyzed the dynamics of citation network and compare it with
the 'rich gets richer' property. We have studied the importance of citation count prediction and analyzed several simple statistical models and
also neural network models for predicting citation count of scientific articles. Citation recommendation is also an important study for new 
researchers to find appropriate published articles for their given research problems. We have differentiated the local and global citation
recommendations based on the context of citation and analyzed several neural network based models. There is also possibility of malicious
citation flow by the authors for self-benefit in the citation network and it is also important to detect those malicious citation flow in the 
dynamics of citation network. We have also studied few effective algorithm to prevent these anomalous citation flow.

The wiki model acts as an open and free knowledge base including large variety of information ranging from history, arts and politics to science and technology and many more fields using wide variety of languages. We have shown our study on the quality of wiki-data based on the article structure, article stability, automatic methods for quality detection and many other features. Apart from that we have also discussed the problems regarding anomaly and vandalism for content editing and retention of reliable editors.

Introduce your topic in \textasciitilde{}2 paragraphs,
\textasciitilde{}750 words.

While Wiley does consider articles on preprint servers (ArXiv, bioRxiv,
psyArXiv, SocArXiv, engrXiv, etc.) for submission to primary research
journals, preprint articles should not be cited in WIREs manuscripts as
review articles should discuss and draw conclusions only from
peer-reviewed research. Remember that original research/unpublished work
should also not be included as it has not yet been peer-reviewed and
could put the work in jeopardy of getting published in the primary
press.

Citations are automatically generated by Authorea. Select~\textbf{cite}
to find and cite bibliographic resources. The citations will
automatically be generated for you in APA format, the style used by most
WIREs titles. If you are writing for~\emph{WIREs Computational Molecular
Science} (WCMS), you will need to use the Vancouver reference and
citation style, so before exporting click Export-\textgreater{} Options
and select a Vancouver export style.

A sample citation:~~\hyperref[csl:1]{(Murphy et al., 2019)}
}
\fi

\section{Wikipedia as a collaborative platform}
Wikipedia models a hypertext collaborative platform along with an open and free knowledge base catering a large variety of information ranging from history, arts, culture, politics to science, technology and many more fields. %has enabled Wikipedia a promising model for collaborative knowledge sharing. 
Being one of the most widely viewed sites (within top ten) in the world since 2007, it spans over 208 languages with a copious amount of articles in each edition. For example, the English-language Wikipedia, the largest in volume, contains more than 6 million articles as of February, 2020. Owing to the collaborative nature and an open-access policy announced by Wikipedia as ``anyone can edit'', a number of challenges have cropped up in maintaining the veracity-quality balance of the content. Although Wikipedia has enforced several rules and strict administrative policies to protect the encyclopedia from malicious activities, a lack of authoritative vigilance prohibits its trustworthiness in academics. In contrast, Wikipedia's structured, complete and detailed evolution history receives increasing attention of the research community in discovering automated solution (e.g. bots, software, API etc.) to meet the goal of quality management.

% Todo Soumya
\subsection{Quality}
The elementary purpose of Wikipedia is free, unbiased, accurate information curation.  To achieve this objective Wikimedia foundation which is the governing body of the platform, has developed labyrinth of guidelines which editors are expected to follow so that highest encyclopedic standards are maintained. These guidelines also enhance accessibility of the Wikipedia articles to a broad community of netizens. We enumerate these guidelines into three categories as discussed below. 
\subsubsection{Article language}
Expressions describing a subject should be neutral. Promotion bearing words such as {\em renowned, visionary, iconic, virtuoso etc.} should not be used. Subject importance should be demonstrated using facts and attribution\footnote{\url{https://en.wikipedia.org/wiki/Wikipedia:Manual\_of\_ Style/Words\_to\_watch}} Prose should have active voice. Jargon needs to be elaborated or substantiated with reference. Any effort to propagate myth or contentious content should be curtailed. An example for this is addition of prefix {\em pseudo} or suffix {\em -gate} which encourages the reader to assume that the subject is factitious or scandalous respectively. Euphemisms (e.g., {\em passed away, collateral damage} and cliches (e.g., { \em lion's share, tip of the iceberg}) disallows presentation of prose directly and hence is restricted. Any unnecessary emphasis in the form of italics, quotations etc. is discouraged. For a complete list of details of content guidelines for the English Wikipedia we refer the reader to the Wikipeida manual of style\footnote{\url{https://en.wikipedia.org/wiki/Wikipedia:Manual\_of\_Style}}.

\subsubsection{Article structure}
These guidelines include proper formatting of the Wikipedia article in terms of section headings, infobox, article name, section organisation etc. The lead section should not be of arbitrary length. The following sections should not be exorbitant in size and bigger sections should be broken into coherent smaller sections. Another requirement is proper positioning of the images with captions and references. In order to alleviate manual labor in improving article structure there have been some automated approaches leveraging advances in machine learning techniques~\cite{jana2018wikiref}.

\subsubsection{Article stability}
These guidelines denote stability of the article, i.e., the respective article should not be subject of frequent \textit{edit wars}. There should not be abusive language exchange among editors and discussions toward improving article quality should organically reach consensus. This is the most difficult objective in collaborative content creation and generally the onus lies in the hands of senior level editors and moderators for smooth conflict arbitration.

\subsubsection{Peer review framework}

Although  Wikipedia has grown significantly in terms of volume and veracity over the last decade, the quality of articles is not uniform~\cite{warncke2015success}. The quality of Wikipedia articles is monitored through a rating system where each article is assigned one of several class indicators. Some of the {\em major} article categories are \textbf{FA}, \textbf{GA}, \textbf{B}, \textbf{C}, \textbf{Start} and \textbf{Stub}. Most complete and dependable content is annotated by an FA ({\em aka featured article}) tag while the lowest quality content is annotated with a Stub tag. The intention behind this elaborate scheme is to notify editors regarding current state of the article and extent of effort needed for escalating to encyclopedic standards\footnote{\url{wiki/Wikipedia:WikiProject Wikipedia/Assessment}}. The editors are expected to rigorously follow the aforementioned guidelines. As has been evident from the guidelines, they are circuitous and often require experience for implementation. Such strict policy adherence have also been sometimes a barrier for onboarding of new editors on Wikipedia which has led to the decline of newcomers over the past decade~\cite{steinmacher2015social,halfaker2011don}. Since it is nontrivial to discern qualifying differences between articles manually, it has given rise to the emergence of automated techniques using machine learning models.

\subsubsection{Computational methods for quality prediction}

Automatic article assessment is one of the key research agendas of the Wikimedia foundation\footnote{\UrlFont{www.mediawiki.org/wiki/ORES}}. One of the preliminary approaches~\cite{halfaker2015artificial} seeking to solve this problem extracted structural features such as presence of infobox, references, level 2 headings etc. as indicators of the article quality.~\cite{dang2016quality} proposed the first application of deep neural networks into quality assessment task where they employed distributional representation of documents~\cite{le2014distributed} without using manual features. The authors in \cite{shen2017hybrid} introduce a hybrid approach, where textual content of the Wikipedia articles are encoded using a BILSTM model. The hidden representation captured by the sequence model is further augmented with handcrafted features and the concatenated feature vector is used for final classification. ~\cite{zhang2018history} is an edit history based approach where every version of an article is represented by $17$ dimensional handcrafted features. Hence, an article with $k$ versions will be represented by $k \times 17$ matrix. This $k$ length sequence is passed through a stacked LSTM for final representation used in classification. ~\cite{shen2019joint} proposed a multimodal information fusion approach where embeddings obtained from both article text as well as html rendering of the article webpage is used for final classification. ~\cite{guda2020nwqm} proposed the first approach which incorporates information from three modes for quality assessment, i.e., article text, article image and article talk page.~\cite{guda2020nwqm} obtains $8\%$ improvement over~\cite{shen2019feature} approach and achieves the SOTA result. A complementary direction of exploration has been put forward by~\cite{li2015automatically,de2015measuring} where correlation between article quality and structural properties of co-editor network and editor-article network has been exploited. An orthogonal direction of research looks into edit level quality prediction which is a fine-grained approach toward article content management~\cite{sarkar2019stre}. 

\subsection{Collaboration among editors}
% Todo Paramita

The workhorse behind the success story of Wikipedia is the large pool of its voluntary editors; an encouragement toward global collaboration influences people to contribute on almost all wikipages. These group of people maintain Wikipedia pages behind the scenes which includes creating
new pages, adding facts and graphics, citing references, keeping the wording and formatting appropriate etc. to lead the articles to the highest level of quality. The achievement of any open collaborative project is hinged on the continued and active participation of its collaborators, and hence, Wikipedia needs to manage its voluntarily contributing editor community carefully. In the days of extreme socio-cultural polarization, algorithmically crafted filter bubbles and fake information represented as facts, editors are highly motivated to contribute to the largest non-biased knowledge sharing platform although their works are not financially compensated most of the times \cite{littlejohn2018becoming}. In these lines there have been several works \cite{yang2016did,muric2019collaboration}, which attempt to understand the dynamics of interaction behaviours of the community in sustaining the health of Wikipedia. 

\subsubsection{Anomaly} 
While investigating the editing behaviours of editors in general context, researchers have found out a taxonomy of semantic intentions behind the edits, and conflicts and controversy are inherent components of the classification. Wikipedia owes its success for several reasons and openness is one of those pillars. Sometimes, the very openness misguides editors to violate Wikipedia's strict guidelines of the neutral-point of view (NPOV), and their disruptive edits cause various kinds of anomalies. We describe the two dominant disputes, produced by the damaging edits as follows.  

\noindent\textit{Vandalism}: With the freedom of editing anything by anyone, Wikipedia has to struggle in stopping the malicious practice of contaminating articles by bad faith edits intentionally. The popular pages like famous celebrities, controversial topics etc. become the frequent targets of vandalism where vandals try to mislead the readers by addition, deletion or modification, which can be termed as hoax. Wikipedia has enforced several strict policies such as blocking and banning Vandals (registered / unregistered editors), patrolling recent changes by adding watch-lists, protecting articles (ex, semi-protected pages) from new editors, random IP addresses etc. In addition to the administrative decisions, bots \footnote{\url{https://en.wikipedia.org/wiki/User:ClueBot\_NG}} are employed to detect and revert the vandalism automatically and finally warn the editors without human intervention. Researchers have proposed various automated ways \cite{Spezzano2019DetectingPT,Kiesel2017SpatioTemporalAO}, i.e., the state-of-the-art techniques based on machine learning \cite{Kumar2015VEWSAW, Susuri2016MachineLB} and deep neural methods~\cite{MartinezRico2019CanDL,Tran2015CrossLanguageLF} in preventing Wikipedia from vandalism. 

\noindent\textit{Edit war}: Apart from the intended malpractice of vandalism, editors often engage themselves in disagreement which further influence them to override each other's contribution instead of dispute resolution. Any such actions violating the three-revert rule\footnote{\url{https://en.wikipedia.org/wiki/Wikipedia:Edit\_warring}} is coined as edit warring in Wikipedia and it promotes a toxic environment in the community. Ultimately in the long-term, the integrity of the encyclopedia will be affected significantly by the damaging effects of edit wars \cite{Ruprechter2020RelatingWA}. Although Wikipedia encourages editors to be \textit{bold}, in contrast a constant refusal to \textit{get the point} is also not entertained.

\subsubsection{Retention of editors}
Historically, Wikipedia managed the numbers of its volunteers quite successfully; however, experts \cite{halfaker2013rise} note that it is at the danger of sharp decline of its active editors due to the lack of the socialization effort. Editors may choose to leave the platform for personal reasons as well as for their disagreement/conflict with their fellow editors. The damage is happening in both ways - when new editors fail to inherit the rules and policies they easily become upset and leave eventually. Experienced editors, on the other hand, can get discouraged because of the continuous upgradation of policies to retain newcomers, or even for the nuisances by the newbies. Two way effort are being taken to combat with this problem -- researchers are coming up with various approaches (see~\cite{Morgan2013TeaAS,Morgan2018EvaluatingTI,yazdanian2019eliciting} and the refereces therein) while Wikipedia itself is running several wikiprojects\footnote{\url{https://en.wikipedia.org/wiki/Wikipedia:WikiProject\_Editor\_Retention}}$^{,}$\footnote{\url{https://en.wikipedia.org/wiki/Wikipedia:Expert\_retention}} to proactively retain its contributors.

%\textcolor{red}{Add one small para on future directions like Souvic}.
\noindent \textbf{Future directions}: Due to the enormous volume of data publicly available from various multilingual wikiprojects, several interesting future directions can be explored. One of the directions is combating repeat abusers who add malicious content annonymously after being blocked through sockpuppetry~\cite{maity2017detection} or collusion. A large volume of effort has been invested in understanding editor behaviour; however, similar large scale exploration need to be done on understanding readers. A promising start in this direction includes the following ~\cite{johnson2020global,ribeiro2020sudden}. However, further work needs to explore the interplay between editors and readers and how these two stakeholders can forge a partnership in mutually beneficial fashion. We refer to the mediawiki research index \footnote{\url{https://meta.wikimedia.org/wiki/Research:Index}} for a comprehensive take on this emerging research scope.

\section{Recent trends in citation dynamics}
% Todo Rima
Research on citation network have always remained essential in solving various problems such as predicting emerging topics, early citation prediction, modelling evolving citation networks. Citation network is a directed graph where nodes could be authors/papers/journals and edges are the citation flows (weighted/unweighted) from one node to another node. %Over the decades, various problems in bibliometric domain are attempted to solve with the help of citation information. %Early years, researchers are involved in finding out the importance of citation in order to identify/early predict some excellent research papers/authors. %Over the years, various metrics (in terms of citations) are proposed to measure the author's contribution to the research.
Using citation networks, one can predict which field/topic could be the `most attractive ones to work on' in the immediate future years. The research dynamics in various fields over the years can be analyzed with the help of the underlying citation network. Various studies uncover the chances of the manifestation of certain new fields/sub-fields by investigating the citation flows from papers of one field to the papers of another field over the time. In recent years, citation count prediction task plays an important role in fund allocations and rewards. Researchers are also interested in building models for automatic citation recommendation while drafting an article. Apart from these, some anomalous practices in exchanging citations have been exposed in the late '90s. Now, such malpractices are becoming more common among the researchers/journals (mostly low ranked). In the rest of this section, we shall discuss each of the above issues in details.

\subsection{Interdisciplinary research in terms of citation interactions}
Various research questions such as ``which field will collaborate with which field in future?", ``Which field will receive more citations from recently published papers?", etc., can be addressed with the help of the underlying citation networks among the articles. Nowadays, research is performed by combining the ideas from multiple disciplines. In~\cite{Hazra:2019} the authors have analyzed the interdisciplinarity among the two basic science fields -- Mathematics and Physics and one fast growing field -- Computer Science. Further they observe how the citation from papers of one discipline flows to the papers of another discipline over the years. They observe that in initial years  huge amount of citation flows from Physics to Mathematics and vice versa. Over the years, Computer Science started gaining citation from Mathematics. In the recent years, both the basic science fields tend to massively cite papers from Computer Science. They observe how popularity of some topics decreases over the time. They found that the Computer Science papers mostly cites the quantum physics sub-field for long time span. In late '90s, Physics mostly cites information theory papers of Computer Science but in recent times it mostly cites papers from machine learning and social \& information networks domain. 

Further, interdisciplinarity has been studied in different fields including biology~\cite{Morillo:2003}, mathematics~\cite{Morillo:2003}, cognitive science~\cite{Till:2016,Kwon:2017}, social science~\cite{Pedersen:2016}, humanities~\cite{Pedersen:2016}. Various studies ~\cite{Barthel:2017,Sayama:2012,Till:2016}
have attempted to propose novel metrics to measure the degree of interdisciplinarity based on researchers’ scientific impact, collaborator’s knowledge, publication history, etc. In addition, metric for measuring interdisciplinarity of an article has been proposed~\cite{Till:2016} where authors’ research area, publications in different domains have been used to define the metric. A study~\cite{Chen:2015} has been carried out to analyze the interdisciplinarity nature of highly cited articles from Thomson Reuters’ databases published in between 1900-2012 years. 

\subsection{Analyzing and modelling the citation dynamics}
Several studies have been carried out in the past to model the temporal dynamics of citation networks. In order to model the temporal dynamics of citation networks, researchers traditionally used preferential attachment~\cite{RevModPhys.74.47} and copying based~\cite{ravikumarFOCS2000} models.

In~\cite{verstak2014shoulders}, the authors investigated the citation behavior of older papers in various fields. They concluded that older articles receive more citations over the years.  It is observed that in 2013, 36\% of citations flowed toward the at least ten years old papers. However, a re-investigation of this study showed that the observations are only partly true since the authors did not take into account the accelerating volume of publications over time. In order to tackle the tug-of-war between obsolescence and entrenchment, 
recently, in~\cite{relaylink}, the authors proposed a complex model based on the idea of \textit{relay-linking} where the older article relays a citation to a recently published article. This model has very less number of parameters and fits with the real data much better than the traditional models. Yet another novel citation growth model called RefOrCite~\cite{PANDEY2020101003} have been proposed recently where the authors allow copying from the references (out-edges) and citations (in-edges) of an article (as opposed to only references in the traditional setup). It is observed that RefOrCite model fits well with real compared to the previous models.

%capture various properties of citation network like rich gets richers, aging. Various standard complex models such as  are employed to understand the dynamics of the citation network. Most of the models mainly focus on the ``rich gets richer’’ (older papers get more citation than newer papers) property to model the network. 

\subsection*{Citation count prediction}
Predicting future impact of scientific articles is important for making decision in fund allocation (by funding agencies), recruitment etc. There are various works~\cite{10.1007/978-3-319-18038-0_51, 10.5555/3060832.3060995, Yuan2018ModelingAP, li-etal-2019-neural} that have been carried out in the past to automatically estimate the citation count of scientific articles. In this article, we shall mainly focus on the recent literature.
In 2015, the authors in~\cite{10.1007/978-3-319-18038-0_51} proposed Trend-based Citation Count Prediction (T-CCP) model where the model first would first learn the type of the citation trends of the articles and then predict the citation count for that trend. All the articles were categorized into five citation trend categories based on the ``burst’’ time (``burst time’’ is the time when the paper gets maximum citations) -- early burst, middle burst, late burst, multi bursts, and no bursts. Two types of features have been used – (a) publication related features like author centric features (i.e., h-index, number of papers published, citation count, number of collaborators), publication venue (average citation count, impact factor etc.) and (b) reinforcement features which are the graph based features (i.e., PageRank, HITS etc.) calculated from  weighted citation network among authors. In their model, they mainly use SVR and SVM (LibLinear) for citation count prediction and classification task respectively. 
In paper~\cite{MayankCIKM2015}, the authors found that the knowledge gathered from citation context within the article could help to predict future citation count. Number of occurrences of the citations for a paper within the article and the average number of words in citation context, have been derived from citation context knowledge. Further they categorized the articles into six citation profiles (PeakInit, PeakMul, PeakLate,  MonDec, MonIncr, Oth) and found that the above two citation context based features are able to nicely distinguish these six categories. In~\cite{MayankJCDL2017}, the authors observed that the long term citation of an article depends on the citations it receives in the early years (within one or two years from its publication date). The authors who cite an article in its early years are called early citer.  Early citers based on whether they are influential or not affect the long term citation count of the article. In most cases, influential authors negatively affect the long term citation of an article.
In~\cite{10.5555/3060832.3060995}, the authors have proposed a novel point process method to predict the citations of individual articles. In their approach they tried to capture two properties -- the ``rich gets richer'' effect and the recency effect. The authors in~\cite{Yuan2018ModelingAP} used four factors -- intrinsic quality (citation count) of a paper, aging effect, Mathew effect and recency effect to derive a model called long term individual level citation count prediction (LT-CCP). In this model they mainly use RNN with LSTM units. It is observed that LT-CPP model achieves better performance than existing models. 
Authors in~\cite{li-etal-2019-neural} proposed a neural model for predicting citation count with the help of peer review text. They mainly learn two deep features -- (a) the abstract-review match mechanism (in order to learn the abstract aware review representation) and (b) the cross review match from peer review text.

%\textcolor{red}{You have not referred to our papers -- CIKM 2015, JCDL 2017 (early citers) etc.? [added]}

\subsection{Citation recommendation}
Often new researchers face difficulties in finding appropriate published research papers while exploring the domain literature and citing published papers. Citation recommendation is a technique that recommends appropriate published articles for the given text/sentence. The sentences present around the reference (placeholder) are called context sentences. Citation recommendation task can be divided into two parts – (i) local citation recommendation, and (ii) global citation recommendation. In case of local citation recommendation, only the context sentences are used. In case of global citation recommendation, the whole article is used as input and the system outputs a list of published papers as output. 
In cite~\cite{bhagavatula-etal-2018-content} the authors proposed a model for the global citation recommendation task where they embedded the textual information (i.e., the title and the abstract) of the candidate citations in a vector space and considered the nearest neighbors as the candidate citations for the target document. Further, re-ranking of the candidate citations was done. They used DBLP (50K articles having an average citation of 5 per article) and PubMed (45K articles with average citation of 17 per article) datasets and also introduced a new dataset OpenCorpus (7 million articles) in the paper. They showed that their model achieved state-of-the-art performance without using metadata (authors, publication venues, keyphrases).  
%Jeong {\em et al.} ~\cite{}%[A context-aware citation recommendation model with BERT and graph convolutional networks | SpringerLink]
In paper~\cite{jeong2019contextaware}, the authors proposed a deep learning model (consists of context encoder and citation encoder) and used a dataset~\cite{10.1145/1772690.1772734} for context aware citation recommendation. Pre-trained BERT~\cite{devlin-etal-2019-bert}  model has been used in order to learn the embedding of the context sentences. GCN has been employed to learn the citation graph embedding from the paper-paper citation graph. They mainly revised two existing datasets -- AAN and FullTextPeerRead (revised version of PeerRead). They showed that their model performed three times better than the SOTA approaches (CACR etc.). The authors in~\cite{10.1145/3383583.3398609} proposed a novel method -- ConvCN -- based on the citation knowledge graph embedding.

\subsection{Detection of anomalous citation flows}
Various anomalous citation patterns have been found to emerge over the years. Various ways of maliciously increasing one’s citation are through {\em self-citations}, {\em citation stacking} among journals, and {\em citation cartel}. Nowadays, authors are more concerned about their position in academia, publication pressure etc. and this leads to most of them adopting unfair means to increase their citation. {\em Citation cartel} is one of the anomalous citation patterns which was first reported in late '90s\footnote{\url{https://science.sciencemag.org/content/286/5437/53}}. {\em Citation cartel} is formed by a group of authors/editors/journals where they cite each other heavily for mutual benefit. The relationship in citation cartel could be author-author, editor-author, journal-journal etc. There are a few cases found where the journal’s impact factor increases rapidly due to this anomalous behavior. {\em Cell Transplantation}\footnote{\url{https://www.cognizantcommunication.com/journal-titles/cell-transplantation}} is a medical journal whose impact factor rapidly increased between 2006 and 2010 (3.48 to 6.20). After investigation carried out by JCR publisher, it was found that one review article published in this journal {\em Medical Science Monitor}\footnote{\url{http://www.medscimonit.com/}} cited almost 91\% papers published in {\em Cell Transplantation} from the time bucket 2008--2009. It was found that the impact factor of the journal  {\em Cell Transplantation} was calculated based on this time bucket\footnote{\url{https://scholarlykitchen.sspnet.org/2012/04/10/emergence-of-a-citation-cartel/}}. Surprisingly, the authors (three out of four) are from the editorial board of this journal. 
In cite~\cite{10.3389/fphy.2016.00049} the authors tried to detect citation cartels. They defined a citation cartel as a group of authors citing each other excessively than they do with other authors' works in the same domain. They observed that there could be multiple reasons like academic pressure, ``publish or perish” concept in academia, fear of losing job, scientific competition etc. behind establishing such citation cartels. It was observed that such unfair means are mostly adopted by low ranked researchers~\cite{Fister2016ANP}. 
%[Fister I Jr, Mlakar U, Brest J, Fister I. A new population-based nature-inspired algorithm every month: is the current era coming to the end? In: StuCoSReC: Proceedings of the 2016 3rd Student Computer Science Research Conference. Koper: University of Primorskapp (2016). p. 33–37]
  In their work, they prepared a multilayer graph where they include paper-paper citation network (directed graph), authors’ collaboration network and authors’ citation networks (weighted directed graph). Finally, citation cartel has been captured from the authors’ citation network. Cartels have been discovered by using Resource Description Framework (RDF) and RDF query language and some threshold has been declared to identify the existence of citation cartel among authors. 
The authors in~\cite{kojaku2020detecting} proposed a novel algorithm -- {\em Citation Donors and REcipients} (CIDRE) to detect the {\em citation cartel} among the journals that cite each other disproportionately to increase the impact factor of the journal. CIDRE algorithm first distinguishes between the normal and malicious citation exchange with the help of few parameters. These parameters are similarity in research areas, citation inflow and outflow. A weighted citation network among 48K journals was constructed from the dataset collected from MAG\footnote{\url{https://www.microsoft.com/en-us/research/project/microsoft-academic-graph/}}. With the help of the algorithm, more than half of the malicious journals were detected (those were actually suspended by Thomson Reuters) in the same year. In addition, CIDRE algorithm detected few malicious journal groups in 2019 whose journals received 30\% of its in-flow citation from the journals in the same group. Such anomalous citations help to grow the impact factor of the journals over the years. In~\cite{mayankICADL}, the authors studied how malicious journals are increasing in the Indian research community and avoiding proper rules and regulations. The analysis has been carried out on Indian publishing group OMICS (considered as predatory by the research community). Surprisingly they observed that such malicious journals share very similar characteristics with various reputed journals.  

%\textcolor{red}{Mayank has a paper on predatory journals. [added]}
\noindent\textbf{Future directions}: In order to gather more citations, malpractices among the journals are rapidly increasing. More research is required to build a mechanism which can automatically predict those (predatory) journals (depending on the topics of the journal). In case of citation recommendation, there is a need for improving the recommendation system such that the system is able to recommend papers that are conceptually similar or exhibit conflicting claims~\cite{F_rber_2020}. Also, prioritizing the citation recommendation would be another help to maintain the page limit given by many conferences~\cite{F_rber_2020}.

%\textcolor{red}{Finally, add one small para on future directions like Souvic [added]}.

% \section*{2. First Level Heading}
\section{Governance}

As noted in the introduction, this section is laid out into two major parts. The former part centers around the spread, automatic detection and containment of hate speech. The latter part deals with bias in media outlets and online recommendation platforms.

\subsection{Hate speech}

\subsubsection{Spread}
The Internet is one of the greatest innovations of mankind which has brought together people from every race, religion, and nationality. Social media sites such as Twitter and Facebook have connected billions of people\footnote{\url{https://techcrunch.com/2018/07/25/facebook-2-5-billion-people}} and allowed them to share their ideas and opinions instantly. That being said, there are several ill consequences as well such as online harassment, trolling, cyber-bullying, and \emph{hate speech}.

\noindent\textbf{The rise of hate speech}: Hate speech has recently received a lot of research attention with several works that focus on detecting hate speech in online social media~\cite{davidsonautomated,del2017hate,Badjatiya:2017:DLH:3041021.3054223,saha2018hateminers,kshirsagar2018predictive}.
Even though several government and social media sites are trying to curb all forms of hate speech, it is still plaguing our society. With hate crimes increasing in several states\footnote{\url{http://www.aaiusa.org/unprecedented_increase_expected_in_upcoming_fbi_hate_crime_report}}, there is an urgent need to have a better understanding of how the users spread hateful posts in online social media. Companies like Facebook have been accused of instigating anti-Muslim mob violence in Sri Lanka that left three people dead\footnote{\url{https://www.theguardian.com/world/2018/mar/14/facebook-accused-by-sri-lanka-of-failing-to-control-hate-speech}} and a United Nations report blamed them for playing a leading role in the possible genocide of the Rohingya community in Myanmar by spreading hate speech\footnote{\url{https://www.reuters.com/investigates/special-report/myanmar-facebook-hate}}. In response to the UN report, Facebook later banned several accounts belonging to Myanmar military officials\footnote{\url{https://www.reuters.com/article/us-myanmar-facebook/facebook-bans-myanmar-army-chief-others-in-unprecedented-move-idUSKCN1LC0R7}} for spreading hate speech. In the recent Pittsburgh synagogue shooting\footnote{\label{pittsburg_shooting} \url{https://en.wikipedia.org/wiki/Pittsburgh_synagogue_shooting}}, the sole suspect, \emph{Robert Gregory Bowers}, maintained an account (@onedingo) on Gab\footref{https://gab.com/} and posted his final message before the shooting\footnote{\url{https://www.independent.co.uk/news/world/americas/pittsburgh-synagogue-shooter-gab-robert-bowers-final-posts-online-comments-a8605721.html}}. Inspection of his Gab account shows months of anti-semitic and racist posts that were endorsed by a lot of users on Gab.
 
\noindent\textbf{Understanding the spread of hate speech}: We perform the first study which looks into the diffusion dynamics of the posts by hateful users in Gab~\cite{Mathew2019}. We choose Gab for all our analysis. This choice is primarily motivated by the nature of Gab, which allows users to post content that may be hateful in nature without any fear of repercussion. This provides an unique opportunity to study how the hateful content would spread in the online medium, if there were no restrictions. To this end, we crawl the Gab platform and acquire 21M posts by 341K users over a period of 20 months (October, 2016 to June, 2018). Our analysis reveals that the posts by hateful users tend to spread faster, farther, and wider as compared to normal users. We find that the hate users in our dataset (which constitute 0.67\% of the total number of users) are very densely connected and are responsible for 26.80\% of posts generated in Gab.

We also study the temporal effect of hate speech on the users and the platform as well~\cite{binny_temporalCSCW}. To understand the temporal characteristics, we needed data from consecutive time points in Gab. As a first step, using a heuristic~\cite{meeder2011we}, we generate successive graphs which capture the different time snapshots of Gab at one month intervals. Then, using the DeGroot model~\cite{degroot1974reaching}, we assign a hate intensity score to every user in the temporal snapshot and categorize them based on their degrees of hate. We then perform several \textit{linguistic} and \textit{network} studies on these users across the different time snapshots. We find that the amount of hate speech in Gab is consistently increasing. This is true for the new users joining as well. We further find that the recently joining new users take much less time to turn hateful as compared to those that joined at earlier time periods. In addition, the fraction of users becoming hateful is increasing as well. Also, we find that the language used by the community as a whole is becoming more correlated with that of the hateful users as compared to the non-hateful ones. The hateful users also seem to be playing a pivotal role from the network point of view. 

\subsubsection{Detection}
% Todo Mithun
Due to the massive scale of online social media, methods that automatically detect hate speech are required. In this section, we explore a few of the methods that try to automatically detect hate speech. Some of the approaches utilize Keyword-based techniques, machine learning models, etc. Perceiving the right features for a classification problem can be one of the challenging tasks when using machine learning. Though surface-level features, such as bag of words, uni-grams, larger n-grams, etc.~\cite{chen2012,vanhee2015} have been used for this problem, since hate speech detection is usually applied on small pieces of text, one may face a data sparsity problem. Lately, neural network based distributed word/paragraph representations, also referred to as \textit{word embeddings/paragraph embeddings}~\cite{Djuric2015} have been proposed. Using large (unlabelled) text corpus, for each word or for a paragraph a vector representation is induced~\cite{mikolov2014} that can eventually be used as classification features, replacing binary features indicating the presence or frequency of particular words. 

Hate speech detection is a task that cannot always be solved by using only lexicon based features/ word embedding. For instance, \textit{`6 Million Wasn't Enough`} may not be regarded as some form of hate speech when observed in isolation. However, when the context is given that the utterance is directed toward Jewish people who were killed in the holocaust by white supremacists and Neo Nazis\footnote{\url{https://www.vice.com/en/article/y3z9aj/a-national-guard-twitch-streamer-said-6-million-wasnt-enough-on-stream}}, one could infer that this is a hate speech against Jews. The above example shows us whether a message is hateful or not can be highly dependent on world knowledge. In~\cite{leigeo2017}, the authors annotated context dependant hate speech and showed that incorporating context information improved the overall performance of the model.

Apart from world knowledge, meta-information (i.e., background information about the user of a post, number of posts by a user, geographical origin) can be used as a feature to improve the hate speech detection task. Since the data commonly comes from the online social media platform, variety of meta-information about the post can be collected while crawling the data. A user who is known to post mostly hateful content, may do so in the future. Existing research~\cite{Mathew2019} has found that, high number of hateful messages  are generated from less number of users.  It has been also observed that men are more likely to spread hate speech than women~\cite{Waseem2016}. Also, the number of profane words in the post history of user has been used as a feature for hate speech classification task~\cite{Dadvar2013}. %From that we can infer that several meta-information can be used as feature to further improve the performance of model.

Nowadays the number of posts which consists of images, audios, and video content are getting shared more in Social media platforms. In~\cite{Gomez2020}, the authors have explored the textual and visual information of images for the hate speech detection task.

 Most of the methods that have been explored earlier were supervised and heavily dependant on the annotated data. Off-the-shelf classifiers such as logistic regression, support vector machines have been extensively used. Recently, deep neural models are being extensively used for the classification task. In~\cite{aluru2020deep}, the authors explored several models such as CNN-GRU, BERT, mBERT for the classification task in 9 languages and observed that with increasing training data, the classifier performance
increased. Another key observation they made was that the relative performance seemed to vary depending on the language and the model. %It can be concluded from that depending on the language and training data size models performance can vary.
While better models for hate speech detection are continually being developed, little study has been done to understand the bias and interpretability aspect of hate speech. In \cite{mathew2020hatexplain}, the authors release a benchmark dataset which could be used to study this. In their work, they show that the models that perform very well in classification do not score high on explainability metrics.

\subsubsection{Counter speech}
After detection of hate speech, we need proper mitigation strategies to stop it from becoming viral. The current methods largely depend on blocking or suspending the users, deleting the tweets etc. This is performed mostly by moderators which is a tedious task for them given the information rate. Many companies like Facebook have started to automate this process but both these methods have the risk of violating free speech. This work~\cite{Katzenbach2020-KATACM-4} identifies various pitfalls with respect to trust, fairness and bias of these algorithms. A more promising direction could be to counter with speech which are popularly known as \textit{counter speech}. Specifically, counter speech is a direct non-hostile response/comment that \textit{counters} the hateful or harmful speech~\cite{richards2000counterspeech}. The idea that `more speech' is a remedy for dangerous speech has been familiar in liberal democratic thought at least since the U.S. Supreme Court Justice Louis Brandeis declared it in 1927. There are several initiatives with the aim of using counter speech to tackle hate speech. For example, UNESCO released a study~\cite{gagliardone2015countering} titled `Countering Online Hate Speech', to help countries deal with this problem.

The frameworks for mitigating hate speech using counter speech involves two school of thoughts. One of them will be to develop fully automatic counter speech generation system, which can output contextually relevant counter speech given a hate speech. Since generating contextual replies to a text is still a nascent area in natural language processing, generating counter speech is expected to be further difficult for AI systems due to the variety of socio-political variables present in them. Hence, a more practical approach could be to find a task force of moderators who can suitably edit system generated counter speech for large scale use. %users who can counter and assist them in generating counter speech, i.e., by generating counter speech which the user can edit. 

One of the earliest computational studies attempted to identify hate and counter users on Twitter and further observe how they interact in online social media~\cite{mathew2020interaction}. The authors created a small dataset of hate-counter reply pairs and observed how different communities responded differently to the hate speech targeting them. The paper further tried to build a machine learning model to classify a user as hate or counter. A followup study~\cite{binnyCounter} on YouTube comments was conducted further to understand how different communities attempted to respond to hate speech. Taking the YouTube videos that contain hateful content toward three target communities: {\sl Jews}, {\sl African-American} (\textit{Blacks}) and {\sl LGBT}, the authors collected user comments to create a dataset which contained counter speech. The dataset is rich since it not only has the counter/non-counter binary classes but also a detailed fine-grained classification of the counter class as described in~\cite{wright-etal-2017-vectors} with a slight modification to the `Tone' Category. The authors observed that the LGBT community usually responded to hate speech via ``humour" whereas the Jew community used messages with a ``positive tone". The work further adds a classification model which can identify a text as counter speech and its type. %While the performance of counter speech was around 0.77 fscore, the type classification was bad due to the scarcity of data. One of the main limitations of this work was the presence of hostile language, which was in later literature was removed from the counter speech types as it may not de-escalalte the situation. While this work can be used to detect whether a speech is counter speech or not, it still cannot be used for counter speech generation purpose, 
The authors in~\cite{qian-etal-2019-benchmark} generated a large scale dataset having hate and their counter replies. They further used this dataset for counter speech generation using seq2seq models and variational autoencoders. One of the limitations of this paper was that the counter speech data annotated through crowd-sourcing were very generic in nature. Hence, a later work~\cite{chung-etal-2019-conan} took help from experts from and NGO to curate a dataset of counter speech toward Islamophobic statements from the social media. While this dataset provides diverse and to the point reply to hate speech, it largely depends on the experts availability. These challenges were compiled in~\cite{tekiroglu2020generating}, where the authors showed how data collection and counter speech generation is dependant on the assistance from the experts. The former paper also highlighted the weakness of the generation models with around 10\% of the automatic responses being proper response to the given hate speech. This reinstates the fact that current generation systems are not capable of understanding the hidden nuances required to generate proper counter speech. 

Another important question that lurks around in the research community is the ``effect of counter speech". While in the case of banning or suspension the effect, i.e., removal of tweet/user is visible, the effect of counter speech is rather subjective in nature. In a recent work~\cite{garland2020countering}, the authors used a classifier to identify ~100,000 hate-counter speech pairs and found reduction of hate speech due to the organised counter speech. While whether this solely was caused by the counter speech or some other reason is still an open question to the research community. 

Overall the counter speech research shows promise but has several unanswered questions about its data collection, execution and effect. Nevertheless, fighting hate speech in this way has some benefits: it is faster, more flexible and responsive, capable of dealing with extremism from anywhere and in any language and it does not form a barrier against the principle of free and open public space for debate. We hope gradually by using counter speech, it will be slowly possible to move toward an online world where differences could exist but not divisions.

%\textcolor{red}{Add one small para on future directions like Souvic} Done.

\begin{figure}[h]
	\centering
	\includegraphics[width=4in]{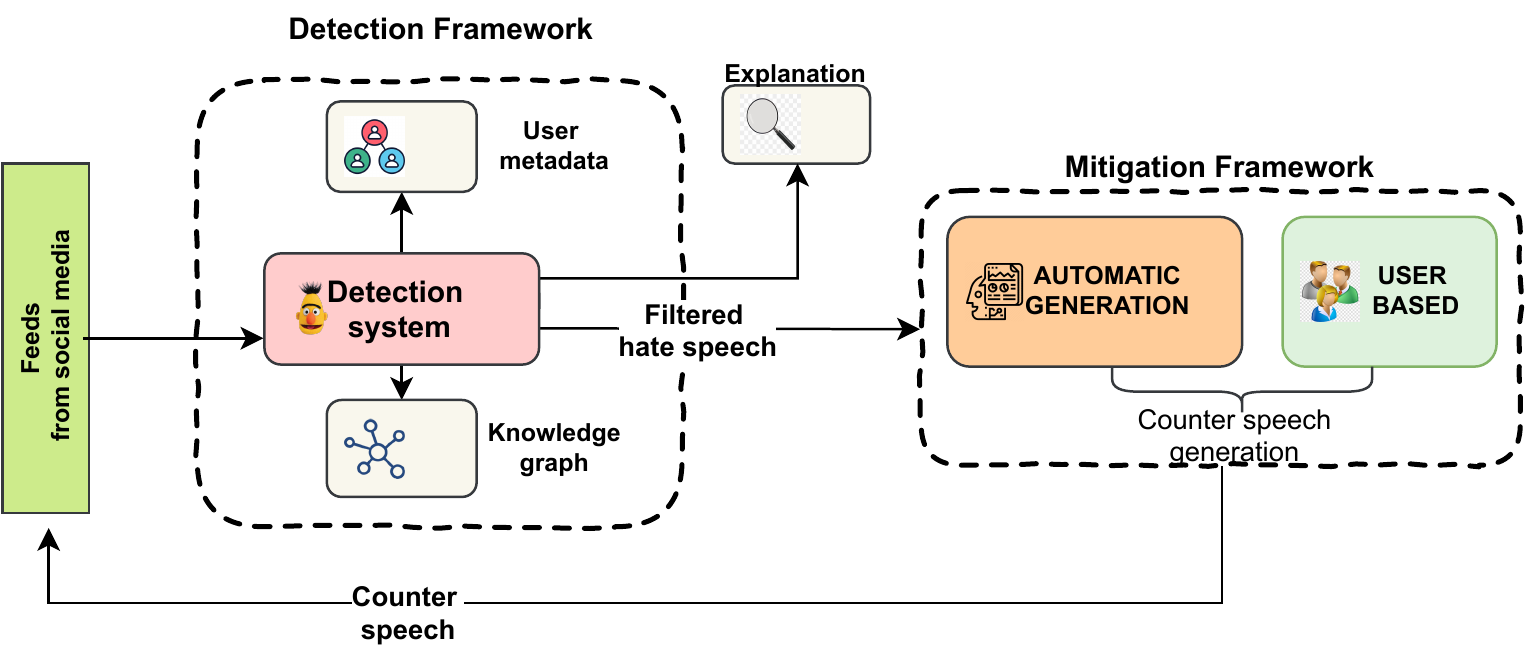}
	\caption{An overview of the hate speech framework.}
	\label{Fig: mitigation_framework}
\end{figure}

\textbf{Future directions: } Increased polarization seems to be spreading hate speech more. Most of the current models have been developed for English language. There is a need for larger and better hate speech datasets for other languages as well. Using transfer learning for improving the task is another direction. Zero or few shot learning would allow models to be able to build models for low resource languages. Finally, an orthogonal but very interesting direction is to understand the complexity of the annotation task itself. Due to the subjective nature of the task, the perception of hate speech is different for people belonging to different demographics. Another important direction is to integrate the detection and counter systems to build an end-to-end framework for effective hate speech detection and countering mechamism, as shown in figure~\ref{Fig: mitigation_framework}.
% Todo Punyajoy

\subsection{Bias and fairness}

As outlined in the introduction, in this section we shall talk about two types of important biases (out of many more that dwells in the online world) -- bias in news reporting and bias in online recommendation systems.
%\textcolor{blue}{I think we need some connecting text here.--AD}
\subsubsection{Media bias}
% Todo Souvic
As we hit 53.6\% Internet penetration worldwide, compounded by an exponential growth in the number of social media users in the developing countries fuelled by cheap data-rates and smartphone-based-accessibility\footnote{\url{https://economictimes.indiatimes.com/tech/internet/internet-users-in-india-to-reach-627-million-in-2019-report/articleshow/68288868.cms?from=mdr}}, the news media continues to play a significant role in shaping political discourse and influencing national priorities. Of late, while on one hand, it has become easier to produce news without adequate references, on the other hand, most newsreaders share news without any verification\footnote{\url{https://www.washingtonpost.com/news/the-intersect/wp/2016/06/16/six-in-10-of-you-will-share-this-link-without-reading-it-according-to-a-new-and-depressing-study/}}~\cite{f3}. In many instances, even the mainstream media houses have been accused of copying and distributing news from other media houses with little or no verification \footnote{\url{http://archiwum.thenews.pl/1/10/Artykul/280476}}. While being informed about news from sources other than direct correspondents is a common practice, distribution of that news without verification indeed is a worrying trend. In many cases, this has led to fake news propagation by the most reputed media moguls. So, quantifying media bias and defining the abstract idea of bias in this context is an important area of research.

\vspace{1mm}
\noindent \textbf{Genesis}: In Manufacturing Consent~\cite{mconsent}, 1988, Edward Herman \& Noam Chomsky saw news media as the propagandist which will find ways to propagate the ``filtered'' message of the rich and powerful to the ordinary masses. Sooner or later in any system, they hypothesize that the news medium will get concentrated in the hands of a few people of power and money and will get manipulated either by ownership or by filtering out news not beneficial for the people in power. Following this model, researchers have hypothesized different kinds of biases like there have been numerous studies examining bias in media especially in the US and the European context. While the term ``bias'' still remains abstract, some studies have put efforts to make a distinction between the computational sense of bias and the journalistic sense of the same making it more scientifically definable and quantifiable.
Journalistic and linguistic studies mostly discuss selection/coverage bias, confirmation/statement bias~\cite{lazaridou2017identifying,lin2011more,nickerson1998confirmation,Saez_Trumper} and psychological/cognitive
biases~\cite{caliskan2017semantics,recasens2013linguistic}. Recently, a lot of work is being done where the researchers are interested to formulate a computational basis for investigating bias. Some works are focused on specific kinds of bias,
such as gender~\cite{Bolukbasi,madaan2018analyze,zhao2017men}, and race~\cite{chouldechova2017fair}. Politics, in particular, is a widely studied and discussed topic. Researchers seek to find ideological political bias of users in social networks~\cite{conover2011predicting,johnson2016identifying,wong2016quantifying}, news media~\cite{baly2018predicting,budak2016fair,laver2003extracting,le2017scalable,ribeiro2018media} and user comments~\cite{ribeiro2017everything,Yigit_Sert}.
D'Alessio and Allen~\cite{d2000media} list three kinds of media bias to be the most widely studied: Coverage/visibility bias~\cite{3Jakob}, gatekeeping bias/selectivity~\cite{6Richard} or selection bias~\cite{7Groeling} (sometimes referred to as agenda bias~\cite{3Jakob}) and statement bias/tonality bias/presentation bias~\cite{3Jakob,7Groeling}.

\vspace{1mm}
\noindent \textbf{Document level bias in reporting}: A news article may choose to cover some aspects of one news and filter other aspects to bias the sentiments of the readers toward/against a specific political party or interest group. Researchers have annotated such sentiment leanings of news articles at the article level~\cite{semevalyo,budak2016fair} or the sentence level~\cite{sent} building document sentiment prediction models based on the annotated data. Following previous researches on sentiment prediction of documents~\cite{semevalyo, zzz, souv1}, dominated by BERT~\cite{bert} based methods, Longformer~\cite{longformer} is shown to be the best choice in the prediction of document level bias.
 
\vspace{1mm}

\noindent \textbf{Media bias is topic \& demographics dependent}: While machine learning models can achieve high accuracy in the task of bias detection, the models are not immune to dataset variations. Inferring media bias is shown to be highly subjective and dependent on the political beliefs of the annotators~\cite{assump}. Also, the demographic shift plays a key role. The authors in~\cite{Cremisini} showed that simple language variations in the data due to change in point of view (a cross-country dataset) can have a devastating effect on the media bias classifier.
In~\cite{assump}, the authors showed that the common belief~\cite{semevalyo} that the bias of the source media house is the bias of any news article originated from it, does not hold well. Also, media houses tend to show different political bias both by magnitude and polarity for different topics (e.g., ABC News and CNN show heavy liberal bias on Gun issues but exhibit conservative bias on immigration issues)~\cite{Cremisini}.

\vspace{1mm}
\noindent \textbf{Source level bias}: Predicting the topical/political bias of individual news outlets is as critical to media profiling as determining factuality. With the advent of user-generated content and exponential rise of digital media, scaling the process of determining media-bias and factuality of reporting of the media houses has become more and more important as everybody who shares an article or screenshot of the same of any source is a news provider now.
While measuring the factuality or bias of each news media is a hard task and requires world-knowledge, the prediction of aggregate factuality or bias of a news media house is relatively straight-forward. Political bias and factuality of reporting have a linguistic aspect (what was written) along with a social context (who read it).  So, the authors in~\cite{classmedia} crawled relevant data from Twitter, Facebook, YouTube \& Wikipedia and studied the impact of different metadata extracted from these sources while classifying the media sources. The evaluation results showed that what was written matters most, and that putting all information sources together yields huge improvements over the current state-of-the-art.
On the other hand, in ~\cite{ribeiro2018media}, the authors studied the demographics of the US population interested in a media source with a specific political inclination, using the Facebook \textit{AdSense} tool, to understand the leanings of the audience of the news media houses and reports high accuracy by doing just that. Along with political bias, they were also able to identify the demographic biases in the consumer population of any media house in a zero-shot setting (i.e., using no training data). A demo of their application can be found here\footnote{\url{https://twitter-app.mpi-sws.org/media-bias-monitor/}}.
TIMME~\cite{timme} supervises a special form of GCNs to identify the political bias of each Twitter user on annotated data and is able to identify the geographic distribution of Twitter users with particular bias which correlates well with the voting pattern of American citizens. They were able to use the same algorithm to identify the bias of each news media house by gathering their Twitter data.

\vspace{1mm}
\noindent \textbf{Future directions}: News media is widely cited as the fourth pillar of democracy. While the health of a democratic institution depends heavily on fair coverage of the institution by the media houses, studies on how computational media bias is related to the health of democracies are lacking. Again, most of the studies in media bias are concentrated on two-party systems and is done on American and European demographics for the English language media while other democracies also face the same problem deserving similar attention. Further research is needed to understand the challenges faced in a multi-party system for other languages in different demographics. Also, event space in media changes very rapidly. So, research in an online learning setup is needed to further enhance the media bias prediction accuracy over time.

%\subsubsection{Fairness in recommendation}
% Todo Abhisek
\subsubsection{Fairness in recommendation}
%\textcolor{blue}{Suggestions have been incorporated in this section.-AD}
The digital platform is full of choices. To help users make intelligent choices, different information filtering systems are deployed in online platforms. Recommendation systems (RSs) are one such omnipresent module. From consumers' perspective, RSs help in finding useful contents; while from platforms' and producers' perspectives RSs bring revenue and profit~\cite{gomez2016netflix, sharma2015estimating}. Given the multi-stakeholder setup and importance of RSs in the livelihood of many of the stakeholders, recommendation fairness is of utmost importance. 

\vspace{1mm}
\noindent \textbf{Filter bubble and evolution of fairness in RSs}: Traditionally, RSs like other information filtering systems are keyed to relevance~\cite{linden2003amazon, smith2017two}. However, over dependency on relevance has led to differential services to different users or different user groups. To describe these effects succinctly, Eli Pariser coined a term `\textit{filter bubble}'~\cite{pariser2011filter}. 
The filter bubble problem is a concern that personalization technologies, including RSs, narrow and bias the topics of information provided to people while they do not notice these facts. To account for these effects, the field of fairness in recommendation first evolved as `Information Neutrality in RSs'. 
Focusing on customer fairness (information neutrality toward customers) Kamishima et. al. \cite{kamishima2012enhancement,kamishima2011fairness} tried
to solve the unfairness issue in RSs by adding a regularization term that enforces
\textit{demographic parity}\footnote{$U_{par}=|E_g[y]-E_{\neg g}[y]|$ can be an instantiation of such demographic parity based regularizers~\cite{yao2017beyond}.}. Such objectives penalized the differences among the average predicted ratings of
user groups (based on sensitive attributes, e.g., gender). However, demographic parity is only appropriate when preferences are unrelated to
the sensitive features. In tasks such as recommendation, user preferences are indeed influenced by
sensitive features such as gender, race, and age~\cite{chausson2010watches,daymont1984job}. Taking a leaf out of the progresses in fairness literature in supervised machine learning~\cite{hardt2016equality, mehrabi2019survey}, Yao et. al.~\cite{yao2017beyond} put forward fairness notions to bridge the gap. They formulated different customer fairness metric by taking a leaf out of the evolution of fairness in supervised learning~\cite{hardt2016equality} and showed their effectiveness in improving customer fairness in recommendation~\cite{yao2017beyond}.
Following their footsteps a number of works focused on `group fairness' in personalized recommendations~\cite{zhu2018fairness,edizel2019fairecsys} where first they quantified biases due to recommendation algorithms toward socially salient groups and proposed methodologies to mitigate such biases. However, a major drawback of many of these works was their negligence toward one of the major stakeholder in RSs, i.e., the producer of items/services. This led to a second school of thought when Burke et. al.~\cite{burke2017multisided, burke2018balanced} first advocated for fairness toward both customers and providers in a recommendation framework. 
Considering RSs as a two-sided affair many nuanced algorithms came into existence considering fairness toward both customers and producers and thus taking a giant step toward a fair marketplace~\cite{mehrotra2018towards,patro2020fairrec,geyik2019fairness}. 

\begin{figure}[h]
	\centering
	\includegraphics[width=4in]{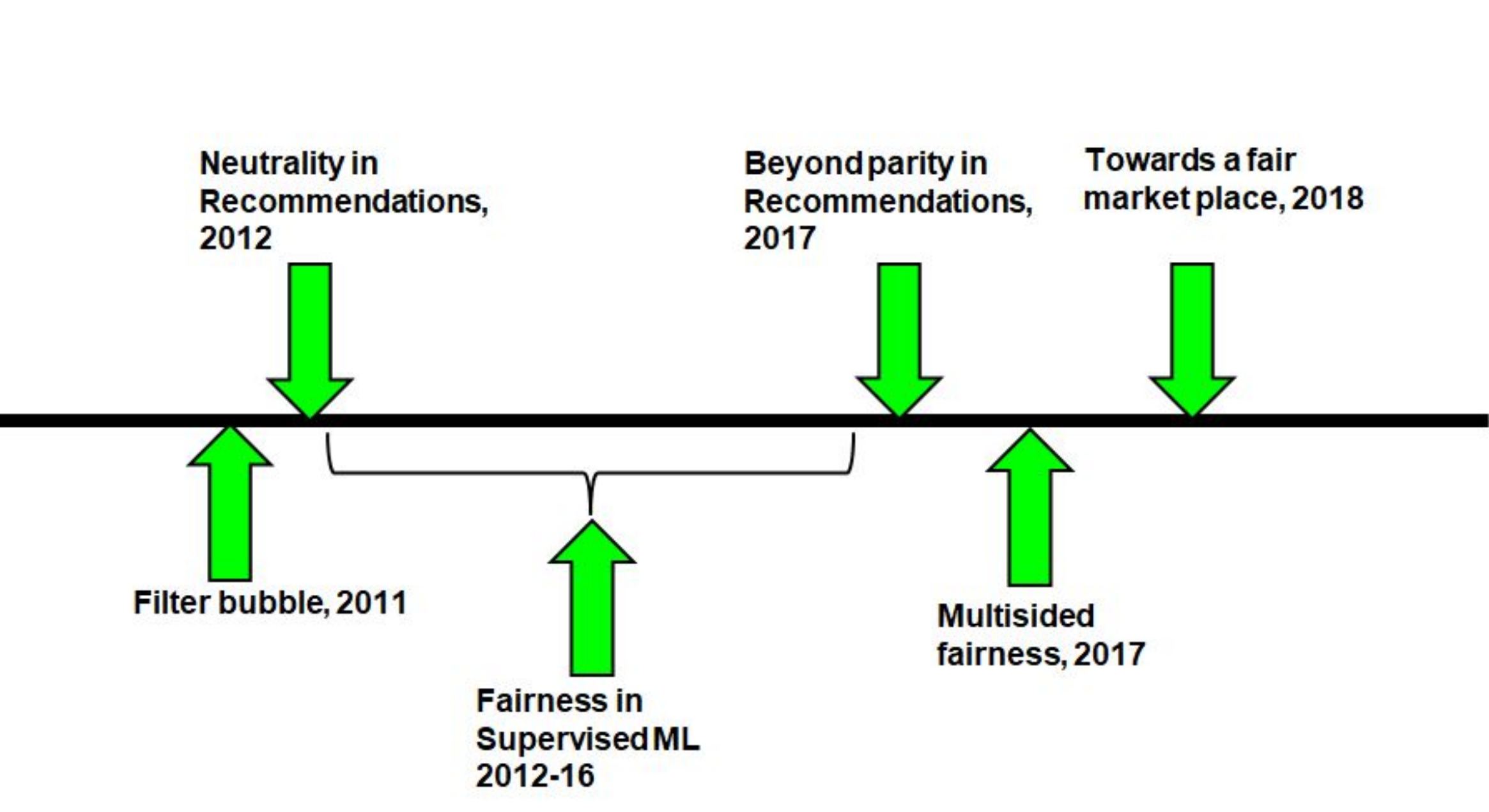}
	\caption{The figure shows the genesis of notions of fairness in recommendation frameworks. In early 2010, it started with the notion of `neutrality in recommendation and by the end of the decade it has evolved into multi-sided fairness toward building a fair marketplace.}
	\label{Fig: FairRecsys}
\end{figure}

\vspace{1mm}
\noindent \textbf{Auditing RSs}: While the fairness community seems to have covered different forms of biases, there is a lack of understanding of the existing online recommendation systems and biases thereof. Understanding of these systems are especially important today due to the emergence of different private label
products (and in-house products) in e-commerce (and OTT) platforms~\cite{Barrett2016Wired,faherty2017amazon,Amazon2019Online}. A private label product is often produced and sold under the
retailer's brand name, providing enough monetary incentive to the platforms to be discriminative against several
other products (or producers) on the platform.
Note that no third party (3P) regulator can quantify such biases because of the lack of access to the exact underlying algorithms and the exact user-item interaction details. To enable such 3P audits, in one of our works, we presented a
novel network-based technique that enabled us to extract important parameters for auditing RSs by considering them as black-boxes~\cite{dash2019network}. With detailed analysis on three different existing online RSs, we first proposed ways to quantify their induced diversity and extent of information segregation~\cite{dash2019network}.
The usefulness of such a framework is manifold: (a)~it sheds light on how recommendations are formed between items based on different item-centric properties, (b)~it can be used as a tool for quantifying and auditing for different consumer-focused metrics e.g., relevance, diversity, information-segregation etc., (c)~it can be used to quantify different biases toward the associated producers of different items/services, and (d)~finally, it also can be used as a tool for mitigating the existing biases by re-wiring few of the recommendation edges.
In another follow-up work, we analyzed the sponsored recommendation on e-commerce marketplaces and showed how they are utilized as a trojan-horse to improve sales of products having special relationship with the marketplace (e.g., private label products)~\cite{dash2021When}. Given that sponsored recommendations potentially have delayed impact on their organic counterparts (by affecting the sales and visibility of products), such studies open up a completely new avenue of research. 

\noindent
\textbf{Future directions: }While the fairness community seems to have covered different forms of biases in recommendation frameworks, it has overlooked the special relationships that may exist between the digital marketplace and a subset of stakeholders, and the biases thereof. Hence, studies of unfairness discovery and mitigation considering the special relationships of platforms remain an under-explored avenue of research till date. The introduction of sponsored search and recommendations complicates the scenario even further. Policies that allow sponsored results to deviate from organic results; while adhering to fairness of marketplace can be another interesting broad direction for further research.

\if{0}\section*{Figures and Tables}

{\label{432859}}

Figures and tables should be numbered separately, in the order in which
they appear in the manuscript. Please embed them in the correct places
in the text to facilitate peer review. Permission to \textbf{reuse or
adapt} previously published materials must be submitted before article
acceptance.
\href{http://wires.wiley.com/go/forauthors\#Resources}{Resources}

Production quality figure files with captions are still to be submitted
separately.
\href{https://authorservices.wiley.com/asset/photos/electronic_artwork_guidelines.pdf}{Figure
preparation and formatting}

\textbf{Captions should stand alone} and be informative outside of the
context of the article. This will help educators who may want to use a
PowerPoint slide of your figure. Explain any abbreviations or symbols
that appear in the figure and make sure to include \textbf{credit lines}
for any previously published materials.\fi

\section{Conclusion}
%\textcolor{red}{@Sayantan In Progress}.
{\label{880788}}
In this survey we have presented a critical rundown on the evolution of the \textit{online infosphere} by depicting some of the research areas that are becoming very crucial at current times. We started our discussion with a view of the infosphere as a collaborative platform, with a dedicated focus on Wikipedia. Wikipedia, the freely available and one of the largest knowledge base, containing a wide variety of information has been a primary focus of an extensive research so far. In this survey we have presented a detailed account of the works on article quality monitoring, editor behaviour and their retention and malicious activities like vandalism.

%In the survey we have first analyzed the data quality of Wikipedia based on several important features of an article construction and it's quality monitoring system. Our study showed that the strict adherence of the quality guidelines of Wikipedia is sometimes becoming a major obstacle for the new editors to contribute. We have concluded the data quality section by putting forward several recent researches about automatic process for quality prediction. The next part of the discussion includes the major problems Wikipedia has to suffer in terms of \emph{Vandalism} and \emph{Edit war} among the collaborators and the strategies that have been employed to tackle those. It has been observed that although Wikipedia encourages the editors to be audacious about their content but it does not accept the stubbornness of sticking to a viewpoint long after getting rejection from the community. Finally we have expressed the issue of decreasing the number of active collaborators and depicted the recent studies to combat it.  

In the next section we have detailed the growth of the citations and collaborations within and across various scientific disciplines that have their roots in the infosphere. In fact, this has resulted in the birth of many new interdisciplinary landscapes. We also discussed how machine learning algorithms can be used to predict future citations as well as for recommending citations. Finally we touched upon various issues related to anomalous citation flows and their behaviour.

%In the beginning of our study we have showcased how the interdisciplinary researches have been flourished over the years. It has been observed that the papers from machine learning and social \& information networks domains dominate the citation in the field of Computer Science in recent years. Some of the recent studies have proposed novel metrics to analyze the interdisciplinarity of a research articles. Coming to the next step of the discussion, we have carried out research on the temporal dynamics of citation networks by exploring the brief idea about recently proposed model \emph{relay-linking} and \emph{RefOrCite}. Apart from that we have discussed how the infosphere has been influencing for predicting future impact of scientific articles and recommending citations for new research by introducing the most recent studies. In the last segment of this study we have shown one of major issue regarding the malicious use of citation flow and explored the scope of research concerning the anomalous citation flow. 

Finally, we summarised the research drives for patrolling the infosphere to suppress the rising volume of harmful content. The discussion started with analyzing the concept of hate speech and its growth over the past few years through the online social media platforms and the adverse impact, thereby, on both the online and the real world. We have shown the massive efforts the research community have put forward in detection and mitigation of such hateful behaviour. However, a lot of issues still remain as open problems and need immediate attention. We observed dense connectivity in the hateful users network by crawling Gab platform  and analyzing their data and found that a significant amount of posts are generated by these hateful users in social media platforms. Additionally, we studied the temporal effect of hate speech on the users by using the Gab data and found the increasing rate of hateful users in the social network. After skimming through the recent literature we have pointed out that by incorporating knowledge based context information for a given post improves the overall performance of hatespeech detection rather than by analyzing only the textual information. Apart from that we have observed that by adding the user information of a given post can be further analyzed to improve the hatespeech detection task. The last segment of the discussion dealt with the bias and discrimination that are becoming pervasive across different online environments like recommendation and news media platforms. %and the recommendation systems. We have unveiled how the recent studies focused on specific categories of bias and shown that the bias detection models are acutely reliant on different demographic dimensions. We then went over the issue regarding the unfairness in recommendation systems in term of \emph{filter bubble} after going through a detailed review of the works done so far addressing this issue and also shown the evolution of fairness studies throughout the past decade.

To conclude, the above mentioned research aspects related to the \emph{online infosphere} have attracted a lot of attention across the board including scientific communities, industry stakeholders and policy-makers. This paper discusses the technical challenges and possible solutions in a direction that utilizes the immense power of AI for solving real world problems but also considers the societal implications of these solutions. As a final note, we could see that the problems related to the anomalies in scientific collaborations and citations, hate speech detection and mitigation in social media and the bias and unfairness in news media and recommendation systems have a lot of open ends thus enabling exciting opportunities for future research.

\if{0}Sum up the key conclusions of your review, highlighting the most
promising scientific developments, directions for future research,
applications, etc. The conclusion should be \textasciitilde{}2
paragraphs, \textasciitilde{}750 words total.

\section*{Funding Information}

{\label{974317}}

You will be required to enter your funding information into the
submission system so that we can apply proper IDs to your funders and
help you comply with any funder mandates.

\section*{Research Resources}

{\label{808103}}

List sources of non-monetary support such as supercomputing time at a
recognized facility, special collections or specimens, or access to
equipment or services.
\href{https://orcid.org/organizations/research-orgs/resources}{Research
Resources} can also be added to ORCiD profiles. For biomedical
researchers, the \href{https://scicrunch.org/resources}{Resource
Identification Portal} supports NIH's new guidelines for Rigor and
Transparency in biomedical publications.

\section*{Acknowledgments}

{\label{749861}}

List contributions from individuals who do not meet the criteria for
authorship (for example, to recognize people who provided technical
help, collation of data, writing assistance, acquisition of funding, or
a department chairperson who provided general support), \textbf{with
permission} from the individual. Thanks to anonymous reviewers are not
appropriate.

\section*{Notes}

{\label{390481}}

Authors writing from a humanities or social sciences perspective may use
notes if a \textbf{comment or additional information} is needed to
expand on a citation. (Notes only containing citations should be
converted to references. Conversely, any references containing comments,
such as ``For an excellent summary of\ldots{},'' should be converted to
notes.) Notes should be indicated by \textbf{superscript letters}, both
in the text and in the notes list. Citations within notes should be
included in the reference section, as indicated below.

\section*{Further Reading}

{\label{153582}}

For readers who may want more information on concepts in your article,
provide full references and/or links to additional recommended resources
(books, articles, websites, videos, datasets, etc.) that are not
included in the reference section. Please do not include links to
non-academic sites, such as Wikipedia, or to impermanent websites.

\section*{\texorpdfstring{{Note About
References}}{Note About References}}

{\label{514168}}

References are automatically generated by Authorea.
Select~\textbf{cite~}to find and cite bibliographic resources. The
bibliography will automatically be generated for you in APA format, the
style used by most WIREs titles. If you are writing for~\emph{WIREs
Computational Molecular Science}~ (WCMS), you will need to use
the~Vancouver reference style, so before exporting click
Export-\textgreater{} Options and select a Vancouver export style.~
\fi

\selectlanguage{english}
\FloatBarrier
%\section*{References}\sloppy
\bibliographystyle{acm}
\bibliography{main,MediaBias/MediaBias}
%\bibliography{/MediaBias}

\phantomsection
%\label{csl:1}{Understanding institutions for water allocation and exchange: Insights from dynamic agent-based modeling}. (2019). \textit{Wiley Interdisciplinary Reviews: Water}, \textit{6}(6). \url{https://doi.org/10.1002/wat2.1384}
\end{document}